\documentclass[10pt,journal,compsoc]{IEEEtran}

\usepackage[misc]{ifsym}

\usepackage{graphicx}
\usepackage{epstopdf}
\usepackage{cellspace, makecell}
\usepackage[many]{tcolorbox}
\usepackage{svg}

\usepackage{mathtools,amssymb,latexsym,amsfonts,stmaryrd}
\usepackage{pbox}
\usepackage{xfrac}
\usepackage{booktabs}
\usepackage{xcolor}
\usepackage{threeparttable}
\usepackage{amsthm}
\usepackage{hyperref}
\usepackage{multirow}
\usepackage{tikz}
\usepackage{mathrsfs}
\usepackage{mathpartir}
\usepackage[ruled,linesnumbered]{algorithm2e}
\usepackage{url}
\usepackage{syntax}
\usepackage{framed}
\usepackage[noend]{algpseudocode}
\usepackage{flushend}
\usepackage{listings}
\usepackage{moresize}
\usepackage{wrapfig}

\usepackage{seqsplit}
\usepackage{alltt}
\usepackage{xspace}
\usepackage{enumitem}
\usepackage{pifont}% http://ctan.org/pkg/pifont
\usepackage{multirow}
\usepackage{amsthm,amsmath,amssymb,lipsum}
\usepackage{mathrsfs}
\usepackage{graphicx}
\usepackage[normalem]{ulem}
\useunder{\uline}{\ul}{}

\usepackage[utf8]{inputenc}
\DeclareUnicodeCharacter{FF1A}{:}

\DeclareMathAlphabet{\mathcal}{OMS}{cmsy}{m}{n}

\usepackage{thmtools}

\declaretheoremstyle[spaceabove=\topsep,notefont=\normalfont\itshape]{mystyle}

\newcommand{\revise}[2]{{\color{red}{\ifx&#1&\else- #1\fi}} {\color{ForestGreen}{\ifx&#2&\else+ #2\fi}}}%
\renewcommand{\revise}[2]{#2}%

\usetikzlibrary{arrows,patterns, decorations.pathreplacing}

\newcommand{\F}{Fig.}
\newcommand{\E}{Eq.}
\newcommand{\T}{Table}
\renewcommand{\S}{Sec.}

\newcommand{\ignore}[1]{}

\newcommand{\parh}[1]{\smallskip\noindent\textbf{#1}}

\lstdefinestyle{base}{
  moredelim=**[is][\color{red}]{@}{@},
  escapeinside={<@}{@>}
}

\usepackage{amssymb}% http://ctan.org/pkg/amssymb
\usepackage{pifont}% http://ctan.org/pkg/pifont

\newcommand\DejaVuttfamily{%
  \fontfamily{DejaVuSansMono-TLF}\selectfont }

\lstdefinestyle{base}{
  moredelim=**[is][\color{red}]{@}{@},
  escapeinside={<@}{@>}
}

\lstdefinelanguage
[x64]{Assembler}     % add a "x64" dialect of Assembler
[x86masm]{Assembler} % based on the "x86masm" dialect
{morekeywords={CDQE,CQO,CMPSQ,CMPXCHG16B,JRCXZ,LODSQ,MOVSXD, %
      POPFQ,PUSHFQ,SCASQ,STOSQ,IRETQ,RDTSCP,SWAPGS, %
      rax,rdx,rcx,rbx,rsi,rdi,rsp,rbp, %
      r8,r8d,r8w,r8b,r9,r9d,r9w,r9b}} % etc.

\usepackage{listings}
\usepackage{fancyvrb}
\usepackage{color}
\definecolor{lightgray}{rgb}{.9,.9,.9}
\definecolor{darkgray}{rgb}{.4,.4,.4}
\definecolor{purple}{rgb}{0.65, 0.12, 0.82}
\definecolor{commentgreen}{RGB}{63,127,95}
\definecolor{pptdy}{RGB}{127,96,0}

\colorlet{myPurple}{blue!40!red}
\definecolor{myOrange}{RGB}{255,192,0}

\newcommand{\enc}[1]{$\phi^{*}_{\theta}$}
\newcommand{\dec}[1]{$\psi^{*}_{\theta}$}

\lstdefinelanguage{Solidity}{
  keywords={len,delete,int,void,payable, public, event, contract, typeof, new, true, false, catch, function, return, null, catch, switch, var, if, in, while, do, else, case, break,struct,const,socklen_t,sa_familty_t,char,sockaddr},
  keywordstyle=\color{violet}\bfseries,
  ndkeywords={class, export, boolean, throw, implements, import, this},
  ndkeywordstyle=\color{darkgray}\bfseries,
  identifierstyle=\color{black},
  sensitive=false,
  comment=[l]{//},
  escapeinside={(*@}{@*)},          % if you want to add LaTeX within your code
  morecomment=[s]{/*}{*/},
  commentstyle=\color{commentgreen}\ttfamily,
  stringstyle=\color{red}\ttfamily,
  morestring=[b]',
  morestring=[b]"
}

\newcommand{\rnum}[1]{\uppercase\expandafter{\romannumeral #1\relax}}

\usepackage{xcolor}

\algnewcommand{\LeftComment}[1]{\Statex \(\triangleright\) #1}

\definecolor{pptbrown}{RGB}{132,60,12}
\definecolor{pptgreen}{RGB}{169,209,142}
\definecolor{pptyellow}{RGB}{255,192,0}

\let\OLDthebibliography\thebibliography
\renewcommand\thebibliography[1]{
  \OLDthebibliography{#1}
  \setlength{\parskip}{0pt}
  \setlength{\itemsep}{0pt plus 0.1ex}
}

\definecolor{pptred}{RGB}{176,35,24}
\definecolor{pptblue}{RGB}{194,214,236}
\definecolor{pptblue1}{RGB}{31,78,121}

\definecolor{pptgreen1}{RGB}{78,173,91}
\definecolor{pptred1}{RGB}{192,0,0}

\definecolor{pptyellow1}{RGB}{203,195,167}
\definecolor{pptgreen2}{RGB}{184,192,176}

\newcommand{\pixel}{\texttt{Contrast}}
\newcommand{\rotate}{\texttt{Rotation}}
\newcommand{\style}{\texttt{Artifying}}
\newcommand{\weather}{\texttt{Snow$_G$}}
\newcommand{\driving}{\texttt{Snow$_D$}}
\newcommand{\face}{\texttt{Eyes}}
\newcommand{\background}{\texttt{Background}}
\newcommand{\clothes}{\texttt{Clothes}}
\newcommand{\legs}{\texttt{Legs}}
\newcommand{\glasses}{\texttt{Glasses}}

\newcommand{\llava}{LLaVA}
\newcommand{\vlm}{CogVLM}
\newcommand{\gpt}{GPT-4V}
\newcommand{\insp}{InstP2P}

\newcommand{\mysubref}[2]{\hyperref[#1]{\ref*{#1}#2}}

\begin{document}

\title{How Multi-Modal LLMs Reshape Visual Deep Learning Testing? A Comprehensive Study
Through the Lens of Image Mutation}

\author{
  {\rm Liwen Wang, Yuanyuan Yuan~$^\text{\Letter}$, Ao Sun, Zongjie Li, Pingchuan Ma, Daoyuan Wu, Shuai Wang~$^\text{\Letter}$\thanks{\Letter~Corresponding authors.}}\\
  The Hong Kong University of Science and Technology\\
  \texttt{\{lwanged, yyuanaq, asunac, zligo, pmaab, daoyuan, shuaiw\}@cse.ust.hk}
}

\IEEEtitleabstractindextext{
\begin{abstract}

    Visual deep learning (VDL) systems have demonstrated their capability to
    understand complex image semantics, paving the way for significant real-world
    applications like image recognition, object detection, and autonomous driving.
    To evaluate the reliability of VDL, a mainstream approach is software testing,
    which requires diverse mutations over image semantics. The rapid development
    of multi-modal large language models (MLLMs) has introduced revolutionary
    image mutation potentials through instruction-driven methods. Users can now
    freely describe desired mutations and let MLLMs generate the mutated images.
    Hence, parallel to large language models' (LLMs) recent success in traditional
    software fuzzing, one may also expect MLLMs to be promising for VDL testing in
    terms of offering unified, diverse, and complex image mutations.
  
    However, the quality and applicability of MLLM-based mutations in VDL testing
    remain largely unexplored. We present the first study, aiming to assess
    MLLMs' adequacy from 1) the semantic \textit{validity} of MLLMs' mutated images,
    2) the \textit{alignment} of MLLMs' mutated images with their text instructions
    (prompts), and 3) the \textit{faithfulness} of how different mutations
    preserve semantics that are ought to remain unchanged.
    With large-scale human studies and quantitative analysis, we identify MLLM's
    promising potential in expanding the covered semantics of image mutations.
    Notably, while SoTA MLLMs (e.g., GPT-4V) fail to support or perform worse in
    editing existing semantics in images (as in traditional mutations like
    rotation), they generate high-quality test inputs using
    ``semantic-replacement'' mutations (e.g., \textit{``dress a dog with
    clothes''}), which bring extra semantics to images and are effective in
    triggering VDL faults; these were infeasible for past approaches. Hence, we
    view MLLM-based mutations as a vital \textit{complement} to traditional
    mutations, and advocate future VDL testing tasks to combine MLLM-based and
    traditional image mutations for comprehensive and reliable testing.

\end{abstract}

}

\maketitle

\IEEEdisplaynontitleabstractindextext
\IEEEpeerreviewmaketitle

\section{Introduction}
\label{sec:intro}

\IEEEPARstart{V}{isual} deep learning (VDL) systems comprehend the semantics of image contents to
enable a wide range of real-world applications (e.g., autonomous driving,
medical imaging diagnosis) that were previously human-reliant. However, VDL
systems can experience drastic failures in their outputs due to their obscure
decision rules, jeopardizing the reliability of VDL systems in safety-critical
applications. Software testing has been very effective in detecting VDL
failures; it mutates input images and defines the accompanying testing oracle
and numeric validation metric. With the metric validating the achieved mutation,
a VDL failure can be detected as a violation of the oracle. In recent years, our
community has proposed a variety of mutation schemes for VDL testing under
different scenarios, such as image classification~\cite{xie2019deephunter},
object detection~\cite{wang2020metamorphic}, image
captioning~\cite{yu2022automated}, and autonomous
driving~\cite{pei2017deepxplore,tian2018deeptest,zhang2018deeproad}.

\smallskip
Inputs of VDL systems are images whose contents have rich semantics. Thus, to
evaluate the reliability of VDL systems via software testing, comprehensively
covering different image semantics is demanding for the input mutation.
Nevertheless, mutating image semantics is fundamentally challenging because the
relation between pixel values and most image semantics is unclear and hard to
model~\cite{yuan2024provably,zhu2020domain,shen2020interpreting,harkonen2020ganspace,ling2021editgan,kim2021exploiting};
most existing methods primarily mutate some pixel-level semantics such as
brightness and contrast~\cite{pei2017deepxplore}.
Some recent works support mutating more complex semantics via representation
learning techniques~\cite{zhu2021low,zhang2018deeproad,shen2020interpreting},
which however require paired images containing the expected mutations and only
apply to specific domains such as driving scenes~\cite{zhang2018deeproad} or
face photos~\cite{shen2020interpreting}.

\smallskip
The recent multi-modal large language models (MLLMs)\footnote{Since MLLMs are
often jointly used with other large models such as Diffusion
model~\cite{ho2020denoising}, CLIP~\cite{radford2021learning},
DALLE~\cite{dalle}, etc., to minimize confusion, we use MLLMs as a general term
for them. We use their own names when referring to their special usage.}, e.g.,
GPT-4V~\cite{achiam2023gpt}, have liberated the freedom of mutations for image
semantics: users can freely describe the anticipated mutation in natural
language and let MLLMs generate the mutated images. This new
\textit{instruction-driven} mutation paradigm is shedding light on (1)
supporting different traditional mutations in a unified prompt-driven manner,
and therefore, one may expect to easily understand, maintain, extend, and
disseminate various input mutation methods in natural language texts. (2)
Moreover, it also shows promising potential in offering a wide range of new
mutations that are hardly achievable by traditional methods, based on the high
natural language comprehension ability of LLMs, and high image synthesis
capabilities of advanced vision backends (e.g., Diffusion
model~\cite{ho2020denoising}).
Thus, one may reasonably anticipate that MLLMs can revolutionize VDL testing, in
parallel to large language model (LLM)'s emerging adoption and success in our
community to boost traditional software
testing~\cite{deng2023large,xia2024universal,GPTScan24}, program
repairing~\cite{peng2024domain,xia2023automated,fan2023automated}, and program
synthesis~\cite{SynthesisWithLLM21,Jigsaw22,AutoSpec24}.

% \smallskip
Nevertheless, unlike traditional methods that explicitly form the mutations
(e.g., via mathematical formulas; see \S~\ref{subsec:explicit}), it is
inherently challenging to interpret how MLLMs generate mutated outputs. Some
recent reports show that MLLMs can generate semantically invalid
images~\cite{du2024stable,chefer2023attend,feng2023trainingfree}, which makes
the follow-up testing meaningless, as those detected ``failures'' are induced by
the VDL system's incorrect usages, not its faults. Thus, it is demanding to
systematically evaluate MLLM's capability in mutating images.

% \smallskip
Following common practice in VDL
testing~\cite{tian2018deeptest,zhang2018deeproad,xie2019deephunter}, we first
measure the quality of MLLM-generated test inputs from: 1$\rangle$ the
\textit{validity} of the overall semantics in mutated images, 2$\rangle$ the
\textit{alignment} of mutated images with text instructions, and 3$\rangle$ the
\textit{faithfulness} of how different mutations preserve semantics that should
be unchanged. This is achieved via a large-scale human evaluation on Amazon
Mechanical Turk~\cite{AMT}, with 20 Ph.D. students experienced in VDL systems
and software testing.
Based on the quality assessment, we then study MLLM's mutation capabilities by
answering two research questions (RQs). \textbf{RQ1}: Can MLLMs unify different
traditional input mutations? \textbf{RQ2}: What benefits can MLLMs bring to
input mutations in VDL testing? Since input validation is crucial to VDL
testing, we further study if existing validation metrics apply to MLLM-based
mutations (\textbf{RQ3}).

% \smallskip
We consider two recently proposed (the only two to our knowledge) pipelines of
leveraging MLLMs for image mutations; both of them implement optimizations for
prompts and image generation modules to provide out-of-the-box usages. Our
evaluations incorporate four SoTA MLLMs (including \gpt) into these pipelines to
achieve 10 representative mutations, and the evaluated datasets cover general image
classification~\cite{deng2009imagenet}, fine-grained dog breed
identification~\cite{dogbreed}, face recognition~\cite{karras2019style}, and
autonomous driving~\cite{cordts2016cityscapes}.

% \smallskip
Our study reveals that MLLM-enabled mutations and traditional mutations are
complementary in VDL testing: MLLMs cannot unify traditional mutations (which
primarily edit the status of semantics that are already in the seed images) and
exhibit low faithfulness and alignment when implementing them, e.g., none of our
evaluated SoTA MLLMs can achieve image rotation. However, MLLMs bring a new
dimension to image mutation: they perform impressively well on
``semantic-replacement'' mutations that replace semantics in the seed images
(e.g., ``\textit{change the background to a library}''); these are uniquely
enabled by MLLMs. Besides, our results show that the two mutation pipelines of
MLLMs may fit distinct scenarios, and call for careful calibrations when
using them. Moreover, we find that existing validation metrics are less
applicable to MLLM-based mutations due to the diversified mutants and
finer-grained operations, and suggest potential improvements.
In sum, this paper makes the following key contributions:

\begin{itemize}%[noitemsep,topsep=0pt,leftmargin=3mm]
      \item This paper conducts the first study on the quality of test inputs
            generated by MLLM for VDL testing, and designs a large-scale human
            evaluation from the validity, alignment, and faithfulness
            perspectives.

      \item Our results reveal the (in-)capabilities of MLLMs in image mutation,
            the limitation of existing validation metrics, and the distinct
            suitability of different mutation pipelines that are designed for MLLMs. We
            also summarize the lessons learned and provide suggestions for future
            improvements.

      \item Our findings better position MLLMs and their enabled
            instruction-driven mutations in VDL testing. While MLLMs cannot edit
            existing semantics in images well like traditional mutations, they
            enable replacing semantics with new ones, being complementary to
            traditional mutations and expanding VDL testing's scope.
\end{itemize}

\section{Background: VDL Testing}
\label{sec:background}

In general, a visual deep learning (VDL) system typically consists of
a visual understanding module and a decision module. The visual understanding
module takes an image as input and extracts features from its content. The
decision module yields outputs (according to the specific VDL application
domain) based on the extracted features. From this perspective, image
classifiers, object detectors, image captioners, and auto-driving systems, etc.,
that were widely tested in our community can all be counted as VDL
systems~\cite{xie2019deephunter,wang2020metamorphic,yu2022automated,pei2017deepxplore,tian2018deeptest}.

\smallskip
The vanilla way of evaluating VDL systems is comparing their outputs with human
annotations over a set of test inputs (i.e., labeled images in standard test
datasets). However, manually annotating test inputs is labor-intensive, impeding
automated and large-scale evaluations. Software testing has achieved remarkable
success in evaluating the reliability of VDL systems. It uses different input
mutation schemes to generate diverse test inputs and forms oracles to enable
automatic ``annotations'' for the test inputs. For example, a slightly rescaled
image (e.g., no less than $50\%$, as determined by the validation metric) should
have the same output as the original image when fed into an object detector.
This way, software testing can identify VDL faults as violations of testing
oracles, where human annotations are no longer required.

\parh{Test Input Generation and Validation.}~Since VDL processes the semantics
of image contents, various mutations and validation metrics have been proposed
to generate legitimate test inputs (see details in \S~\ref{sec:mutation}). In
short, the mutations broadly modify different semantics to explore potential
application scenarios of VDL systems. The numeric validation metrics quantify
the extents of achieved mutations and the validity of test inputs, in order to
control the mutation and ensure meaningful testing results.
Note that different from adversarial perturbations\footnote{Such perturbations
are leveraged by adversarial attacks to generate malicious inputs (i.e.,
adversarial examples) to fool a VDL system~\cite{goodfellow2014explaining}.}
that add invisible noise to images and study the ``worst-case'' attack surfaces
of VDL systems, mutations employed in VDL testing holistically alter images to
diversify image semantics, exploring different usage scenarios of VDL systems.

\parh{Testing Oracle.}~Metamorphic testing (MT) stands out as one mainstream
testing paradigm of VDL testing in recent
years~\cite{wang2020metamorphic,yu2022automated,
tian2018deeptest,xie2019deephunter,yuan2021perception,zhang2018deeproad,yuan2024provably}.
MT checks whether a VDL system's outputs are aligned with the relation (e.g., an
equality relation) defined in an MT testing oracle, when the VDL is processing
an input and its mutated version. Unalignment shows that one of the inputs is
ill-processed and a VDL fault is triggered. Since testing oracles are critical
to detect VDL faults, input mutations themselves should properly preserve the
oracles (e.g., within specific semantic ranges, as determined by the validation
metric); see examples in \S~\ref{sec:mutation}.

\section{Image Mutations and Validation}
\label{sec:mutation}

\begin{figure*}[!ht]
    \centering
    % \vspace{-5pt}
    \includegraphics[width=0.98\linewidth]{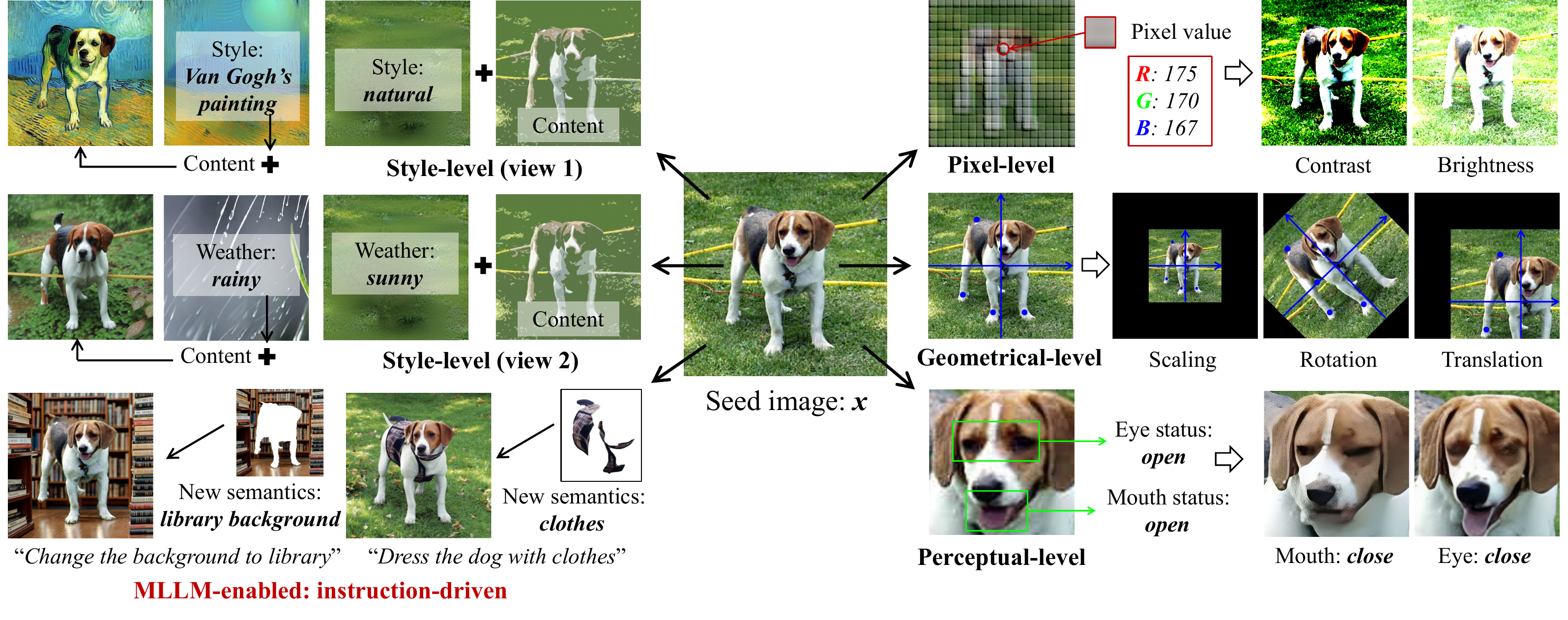}
    \caption{Decomposition of image semantics and their corresponding
        mutation schemes.}
    \label{fig:mutation}
\end{figure*}

\parh{Image Decomposition: A Motivating Example.}~To understand VDL and the
design of input mutations, we first decompose the dog image in
\F~\ref{fig:mutation} into different semantics. Starting from a lower level, the
image is a matrix of pixels, and the magnitude of pixel values determines the
brightness and contrast of the image. Besides, the spatial arrangement of pixels
reflects geometrical properties of the image, such as the scale and the rotation
angle. A reliable VDL system that recognizes dogs should be resilient to changes
of these semantics. At a higher level, the image contains a dog and the
background grass; the dog itself has rich semantics such as the action and the
orientation. Similarly, the semantics in background may vary with the working
scenario of the VDL, e.g., the grass may be replaced by snow in a winter scene.
Overall, the dog-recognition VDL should focus on the semantics of the dog and be
invariant to irrelevant semantics in the background.

\smallskip
Based on the above semantic decomposition, the following image mutation
schemes have been proposed by existing works.

\subsection{Explicit \& Mathematical Transformations}
\label{subsec:explicit}

\parh{Pixel-level.}~Mutations belonging to this category can be formed as
uniformly changing pixel values via $\mathbf{x}$ \texttt{op} $c$, where $c$ is a
scalar value and $\mathbf{x}$ is a matrix denoting the seed image. \texttt{op}
decides the mutated semantics. For example, \texttt{op} is the add operation if
the mutation changes the image brightness, i.e., adding all pixel values in
$\mathbf{x}$ with a positive number $c$ increases $\mathbf{x}$'s brightness.
Similarly, mutating image contrast can be formulated as $\mathbf{x} \times c$,
which enlarges the difference between pixel values in $\mathbf{x}$.
Representative examples are shown in \F~\ref{fig:mutation}.

\parh{Geometrical-level.}~As geometrical properties are determined by the
spatial arrangement of pixels, existing works implement these mutations using
affine transformations, which essentially move pixels to new localizations and
preserve the image's parallelism (i.e., two parallel lines remain parallel in
the mutated image). Formally, suppose pixels of the image $\mathbf{x}$ are in a
rectangular coordinate and the centering pixel is the origin, the $(i, j)$-th
pixel's localization in the mutated image can be calculated as $[i',
j']^\intercal = \mathbf{A} \times [i, j]^\intercal + \mathbf{b}$, where
$\mathbf{A}$ and $\mathbf{b}$ characterize the geometrical mutation. For
example, $(\mathbf{A}_{S}, \mathbf{b}_{S})$ and $(\mathbf{A}_{T},
\mathbf{b}_{T})$, w.r.t. scaling and translation mutations, are shown in
\E~\ref{equ:geo},

\begin{equation}
    % \small
    \label{equ:geo}
    \mathbf{A}_{R} =
    \begin{bmatrix}
        \cos \theta, & -\sin \theta \\
        \sin \theta, & \cos \theta  \\
    \end{bmatrix},
    \mathbf{b}_{R} = \mathbf{0};
    \;
    \mathbf{A}_{T} = \mathbf{0},
    \mathbf{b}_{T} =
    \begin{bmatrix}
        t_x \\
        t_y \\
    \end{bmatrix}
\end{equation}

\noindent where $\theta$ specifies the rotating angle. $t_x$ and $t_y$ indicate
the vertical and horizontal translation distances (see \F~\ref{fig:mutation}).

\parh{Validation and Application Scope.}~Validating the above mutations is
straightforward: developers can directly control the mutations via their
parameters (e.g., $c$ in pixel-level mutations) and ensure generating valid test
inputs by setting reasonable parameter values. These mutations are mostly
adopted in testing safety critical VDL systems like autonomous
driving~\cite{pei2017deepxplore,tian2018deeptest} due to their neat and explicit
formulations. In addition, given these mutations' applicability to almost all
RGB images, they are widely used in testing more general VDL tasks like image
classification~\cite{xie2019deephunter}. 

\subsection{Implicit \& Data Exploration}
\label{subsec:implicit}

Recent works diversify the mutated semantics with data-driven mutations. These
mutations implicitly explore patterns in some images and then apply the patterns
to mutate others.

\parh{Style-level \& Application Scope.}~From a high level, each image can be
decomposed as its style and content, and VDL systems should primarily focus on
the content rather than style. Thus, existing VDL testing works propose
style-level mutations, which generate images of distinct styles but presumably
retain the content of the seed image. The ``style'' has been studied from two
aspects. First, the style can be viewed as describing how the image is produced,
e.g., the seed image in \F~\ref{fig:mutation} is taken in reality and its style
is ``natural''. In that sense, the style can be changed by generating an image's
artistic-stylized variants (e.g., an image having the same dog but painted using
a Van Gogh style, as in \F~\ref{fig:mutation}). These mutations over ``artistic
styles'' have been shown effective in flipping image classifier's outputs in
recent testing works ~\cite{xie2022boosting,yuan2024provably}, and some works
also employ them to reveal VDL system's bias to non-content
semantics~\cite{geirhos2018imagenet}, showing their high value in VDL research.

Besides, some other works decompose the weather condition as the style and
separate it from the main content. The weather conditions are transferred
between images to generate test inputs, as shown in \F~\ref{fig:mutation}.
However, these weather-targeted style-level mutations are specifically designed
for testing VDL-based autonomous driving~\cite{zhang2018deeproad}, and show
limited applications in other testing scenarios.

\parh{Perceptual-level \& Application Scope.}~Recent VDL testing focuses on
fine-grained semantics of the objects in images and implements perceptual-level
mutations~\cite{dunn2021exposing,kang2023deceiving,yuan2024provably}, e.g., in
\F~\ref{fig:mutation}, the dog's eyes and mouth can be mutated to evaluate
whether the VDL accurately recognizes different dogs~\cite{dunn2021exposing}.
Similar techniques can be applied to face photos for testing and improving
finer-grained VDL tasks like face
recognition~\cite{luo2021fa,tran2017disentangled}, face
aging~\cite{liu2019attribute,makhmudkhujaev2021re}, crowd
counting~\cite{wang2019learning,liu2021cross}, etc.

\parh{Validation.}~Both style-level and perceptual-level mutations are
implemented based on generative models like
GANs~\cite{zhu2017unpaired,zhang2018deeproad,karras2019style}. To control the
mutations, existing works propose optimization-based solutions, e.g., the
cycle-consistency mechanism is formed during training
in~\cite{zhu2017unpaired,zhang2018deeproad} to better separate styles from
images. However, such constraints derived from optimizations lack explicit
guarantee, e.g., the cycle-consistency can be violated when transferring styles
among images of dissimilar objects~\cite{Cycleganfail,yuan2024provably}.
Hence, existing testing works additionally adopt numeric metrics, such as
Inception Score (IS)~\cite{salimans2016improved}, Frechet Inception Distance
(FID)~\cite{heusel2017gans}, to measure the validity of generated test inputs.
Since variations on fine-grained semantics in style- and perceptual-level
mutations can hardly be captured via pixel values, recent works propose to
measure the extents of achieved mutations via image embedding\footnote{A
representation of the image that captures its semantics. The term embedding may
be alternatively dubbed as ``feature'' in some
works.}~\cite{missaoui2023semantic,kang2023deceiving,yuan2024provably} (e.g.,
obtaining style and content embeddings, or using embedding models like
CLIP~\cite{radford2021learning,missaoui2023semantic} to compare image semantics
with text descriptions).

\subsection{MLLM-Based Mutations}
\label{subsec:llm}

In line with past image mutations in VDL testing, we explain how MLLMs can
boost the mutation schemes from two aspects.

\smallskip
\parh{Unification \& Generalization: Instruction-Driven.}~MLLMs have
revolutionized image mutation by enabling text-to-image translation. Users can
describe the desired mutation for a seed image using text instructions, and
MLLMs generate the mutated image accordingly.
This instruction-driven paradigm enables VDL developers to uniformly implement
previous mutations introduced in \S~\ref{subsec:explicit} and
\S~\ref{subsec:implicit} and apply domain-specific mutations in
\S~\ref{subsec:implicit} to more general scenarios, by designing the proper text
instruction.

\smallskip
\parh{Expansion: Semantic-Replacement.}~The instruction-driven paradigm also
largely expands the covered semantics in image mutations. Unlike previous
mutations that edit the status of semantics already presented in the seed image,
MLLMs bring new mutations that can replace image semantics with new ones; we
refer to this new category as semantic-replacement in the rest of this paper.
For example, the dog image's main body can be replaced with clothes (see
\F~\ref{fig:mutation}) to further challenge the VDL system's recognition.

\smallskip
\parh{Comparison with Object Insertion.}
Knowledgeable readers may wonder the differences between semantic-replacement
mutations with the object insertion adopted in testing object-based VDL
systems~\cite{wang2020metamorphic,yuan2021perception,yu2022automated}. We
clarify that semantic-replacement mutations operate in semantic-level and are
finer-grained, which often require reasoning the relation and hierarchy between
semantics (either intra- or inter-object). For example, the ``dressing with
clothes'' mutation in \F~\ref{fig:mutation} needs to understand and identify the
dog's body and \textit{replace} it with appropriate clothes. Object insertion,
in contrast, does not perceive image semantics and only relies on object
locations in the image, so that the inserted object will not overlap with
existing ones. Moreover, prior object insertion mutation is not end-to-end,
as they rely on object localization results from object detectors; we thus
omit it in this study.

\section{Research Motivations}
\label{sec:motivation}

\parh{Limitations of Past Methods.}~Although the fast development of input
mutations shows promising results in VDL testing, maintaining and developing
them often requires expertise in different domains. Moreover, the more advanced
style- and perceptual-level mutations show limited application scopes. For
instance, weather-targeted mutations may perform worse in non-driving-scene
images, and the face-related perceptual-level mutations only apply to highly
aligned face photos (as they rely on a unified mask to pinpoint eyes, mouth,
etc.). Some perceptual-level mutations may require manually annotating the
to-be-mutated semantics~\cite{shen2020interpreting,zhu2021low}.

From these perspectives, we envision that MLLMs can offer an
\textit{instruction-driven} mutation paradigm, which is more general, automated,
and highly flexible for VDL testing. MLLMs have shown high capabilities in
natural language understanding and image synthesis, and therefore, various kinds
of mutations that were hardly or tediously achieved by traditional methods may
be easily implemented via MLLMs by directly providing instructions. Moreover,
since various (subtle or complex) mutations are unifiedly described in natural
language (as shown in \F~\ref{fig:mutation}), the mutations can be easily
understood, maintained, calibrated, and extended by VDL developers and the
community.\footnote{In some sense, this new
\textit{instruction-driven} mutation paradigm is comparable to how query-based
software security testing tools, e.g., CodeQL~\cite{codeql}, maintain varying
types of vulnerabilities in a unified query language.}

\parh{Unclear MLLM Internals.}~Despite the optimism, MLLM's image generation is
hard to interpret due to the unclear internals. Per recent reports, MLLMs often
fail to mutate images according to the given instructions~\cite{li2023seed}, and
may generate broken and ill-formed images from time to
time~\cite{du2024stable,chefer2023attend,feng2023trainingfree}. Hence, we
believe our community still lacks understanding of how exactly MLLMs can boost
VDL testing, and a comprehensive study is urgent for VDL developers to
understand MLLM's (in-)capability of mutating images. In addition, considering
that input validation is crucial to ensure meaningful testing results, it is
also demanding to evaluate whether existing validation metrics can still be
applied to MLLM-based mutations.

\section{Research Study Setup}
\label{sec:setup}

Our study answers the following research questions (RQs).

\begin{itemize}
  \item \textbf{RQ1:}~\textit{Can MLLMs unify different traditional input
  mutations? }
We answer this RQ by assessing the quality of 1) how MLLMs re-implement
traditional mutations and 2) how MLLMs extend prior domain-specific mutations to
general scenarios.

\item \textbf{RQ2:}~\textit{What benefits can MLLMs bring to input mutations
  in VDL testing?}
We answer this RQ by 1) assessing the quality of test inputs generated by the
semantic-replacement mutations, and 2) evaluating their effectiveness in
identifying VDL faults.

\item \textbf{RQ3:}~\textit{Are existing input validation metrics reliable
  when facing MLLM-based mutations?}
We answer this RQ by evaluating whether existing validation metrics can reflect
1) the validity of test inputs generated by MLLMs and 2) the varied extents of
different MLLM-based mutations.
\end{itemize}

Our RQs are based on the quality assessment of test inputs generated via
different mutations, which requires perceiving and understanding image
semantics. We therefore conduct a large-scale human study to evaluate image
quality.
Below, we introduce our evaluated aspects for the quality assessment.

\subsection{Evaluated Aspects}
\label{subsec:aspects}

Following design principles in existing input mutations and validation metrics,
we consider the following three aspects.

% \smallskip
\parh{Alignment.}~The achieved mutation must be aligned to the expectation. This
is important as users need to know which mutated semantics trigger VDL faults
and diagnose the VDL's faulty behavior using the testing results. For example,
if a rotation mutation does not rotate the image as expected, users will falsely
interpret the tested VDL system as behaving correctly (\textit{false negatives}).

% \smallskip
\parh{Faithfulness.}~A mutation should faithfully preserve the semantics that
are ought to be unchanged. For example, to close eyes of face photos, the
mutation should let other semantics (e.g., background, hair, nose) remain
unchanged and only close the eyes. Otherwise, the unanticipated mutation may
break the testing oracle and bring \textit{false positives}. The follow-up
diagnosis for VDL's faulty behaviors can also be misled.

% \smallskip
\parh{Validity.}~Besides the alignment and faithfulness that focus on the
mutated and unmutated semantics, we also holistically consider the validity of
all semantics in the whole image. This is crucial as an aligned and faithful
mutation may still generate unrealistic images. For example, as will be shown in
\F~\ref{fig:example}, GPT-4V replaces a portrait's eyes with closed ones
(potentially from other portraits in its ``database'') when asked to close the
eyes. Although the mutation aligns the requirement and other semantics are
untouched, the sharp transition around eyes makes the mutated image unrealistic,
which rarely appear in real-world scenarios.

\subsection{MLLMs and Their Pipelines}
\label{subsec:llm-pipeline}

\begin{figure*}[!ht]
  \centering
  \includegraphics[width=0.98\linewidth]{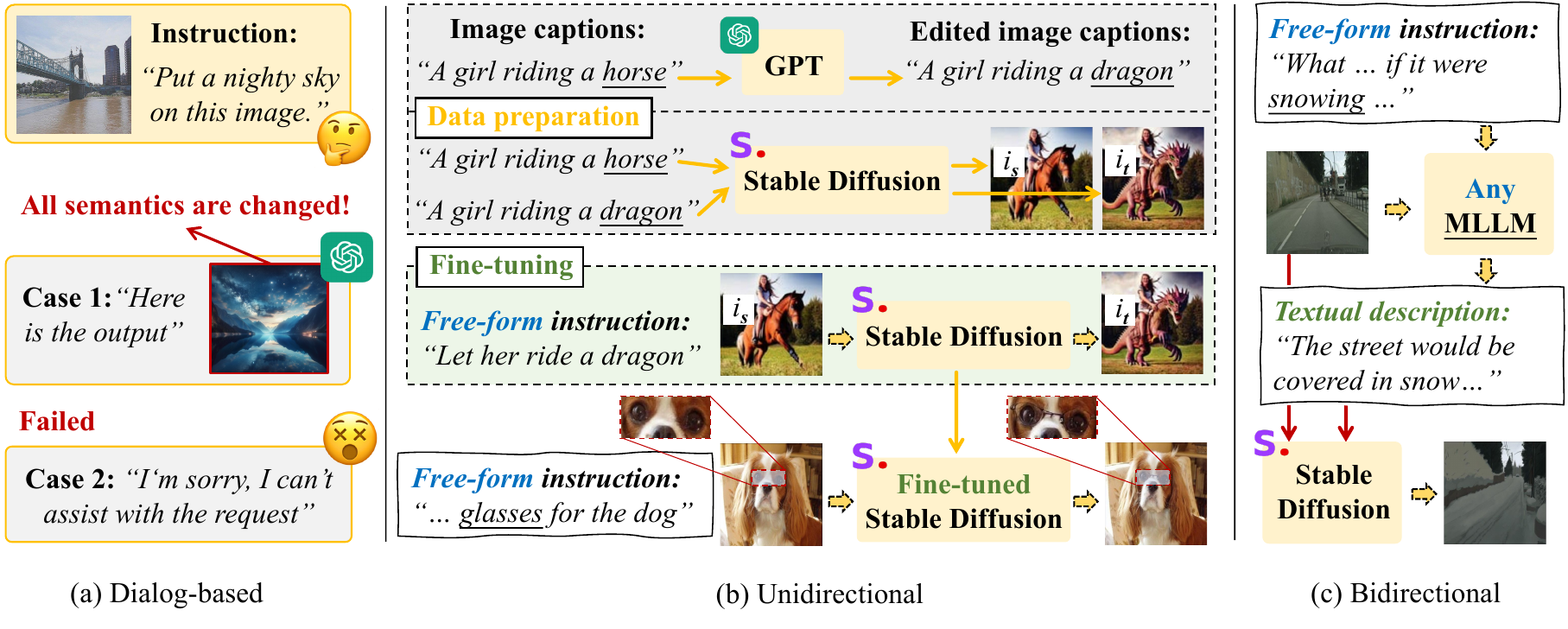}
  \caption{Pipelines of MLLM-based mutations. \F~\mysubref{fig:pipeline}{(a)}
  shows the most straightforward dialog-based pipeline.
  \F~\mysubref{fig:pipeline}{(b)-(c)} illustrates two pipelines specifically
  optimized for image mutations. The unidirectional pipeline shown in
  \F~\mysubref{fig:pipeline}{(b)} requires fine-tuning a MLLM and should be used
  along with this tailored MLLM. The bidirectional pipeline presented in
  \F~\mysubref{fig:pipeline}{(c)} is post-hoc and supports incorporating
  different MLLMs.}
  \label{fig:pipeline}
\end{figure*}

\parh{MLLMs.}~To date, two pipelines are proposed by existing works to employ
MLLMs for image mutations. As illustrated in
\F~\mysubref{fig:pipeline}{(b)-(c)}, these pipelines are optimized from
different angles to avoid issues in prompt engineering or image generation
modules, providing out-of-the-box usages for non-expert in MLLMs.
In particular, the first pipeline (\F~\mysubref{fig:pipeline}{(b)}; see more
details below) relies on a specifically fine-tuned MLLM, we therefore use the
MLLM provided by the authors~\cite{brooks2023instructpix2pix}, along with this
pipeline, as a whole for evaluations and refer to it as \insp. The second
pipeline is post-hoc and supports incorporating different MLLMs; thus, we use
three SoTA and representative MLLMs, including \llava~\cite{liu2024visual},
\vlm~\cite{wang2023cogvlm}, and \gpt~\cite{achiam2023gpt} into this pipeline
during evaluations.

\parh{Dialog-Based.}~The most straightforward way to mutate images is building a
dialog with the MLLM, as shown in \F~\mysubref{fig:pipeline}{(a)}. For example,
the users first provide the seed image and a prompt of the mutation instruction
to \gpt\ and let \gpt\ return the mutated image. However, our preliminary
explorations show that this dialog-based pipeline fails to generate test inputs
in \textit{all} cases we try: \gpt\ either generates an entirely new image
(nearly all semantics are changed) or responds that it cannot generate the
image. Similar issues are also noted
in~\cite{fu2024mgie,huang2023smartedit,huang2024diffusion},

\smallskip
To alleviate this issue, one may expect to use prompt engineering to better
guide the MLLM. Nevertheless, our preliminary study shows that it is hard to
automatically tune MLLM prompts and achieve improved performance constantly on
different images. Moreover, we argue that ``tuning prompts'' is less practical
in our context, as VDL developers (main users of VDL testing tools) are often
not experts in prompt engineering; MLLM should provide out-of-the-box support
for implementing mutations. In this regard, recent works propose to equip MLLMs
with Diffusion model~\cite{rombach2022high}, a more sophisticated image generation
model, and design two representative optimization pipelines to orchestrate their
specialties in cross-modal translation and image
generation~\cite{fu2024mgie,huang2023smartedit,chen2023llava,caffagni2024r,huang2024diffusion}.

\parh{Unidirectional \& Fine-tuning-based.}~As shown in the bottom of
\F~\mysubref{fig:pipeline}{(b)}, \insp\ designs an unidirectional pipeline which
directly uses a free-form text instruction and the seed image to generate the
mutated image. In general, since the text instruction primarily describes the
mutation and does not include details in the images, \insp\ requires
fine-tuning a Diffusion model to support diverse mutations. 

Specifically, as illustrated in the top and middle of
\F~\mysubref{fig:pipeline}{(b)}, a GPT model is first employed to edit a given
textual caption to obtain a new caption. Then, the two captions are fed into a
pretrained a Stable Diffusion model~\cite{rombach2022high} to generate two
images, $i_s$ and $i_t$, respectively.
Finally, a free-form textual instruction is generated using GPT to briefly
summarize the changes between the two captions; this textual instruction,
together with the two generated images $i_s$ and $i_t$, forms a data point for
fine-tuning the Diffusion model, and the Diffusion model is tuned to generate
$i_s$ when given $i_t$ and the textual instruction.
After fine-tuning, the Diffusion model is able to generate a mutated image for
any unseen images and textual instructions. We refer interested readers
to~\cite{brooks2023instructpix2pix} for the fine-tuning details.

\parh{Bidirectional \& Post-hoc.}~Another pipeline, as designed
in~\cite{fu2024mgie,huang2023smartedit,chen2023llava,caffagni2024r,huang2024diffusion},
leverages MLLMs to transform the mutation instruction based on the seed image,
such that details in the seed image can be included in the transformed
instruction. As shown in \F~\mysubref{fig:pipeline}{(c)}, this pipeline is
categorized as bidirectional, since it implements the workflow of image
$\rightarrow$ text $\rightarrow$ image. Specifically, due to MLLM's promising
capability of understanding images, users first feed the seed image to the MLLM
and let the MLLM describe what the image would look like after applying a
mutation. Here, the MLLM's output is a textual description of semantic-level
details in the mutated image. 
This finer-grained text description, together with the seed image, is then fed
to the Diffusion model to generate the mutated image. The superiority of this
pipeline (compared with the dialog-based one) has been empirically justified
in~\cite{fu2024mgie,huang2023smartedit,chen2023llava,huang2024diffusion}.

\begin{figure*}[!ht]
  \centering
  \includegraphics[width=0.95\linewidth]{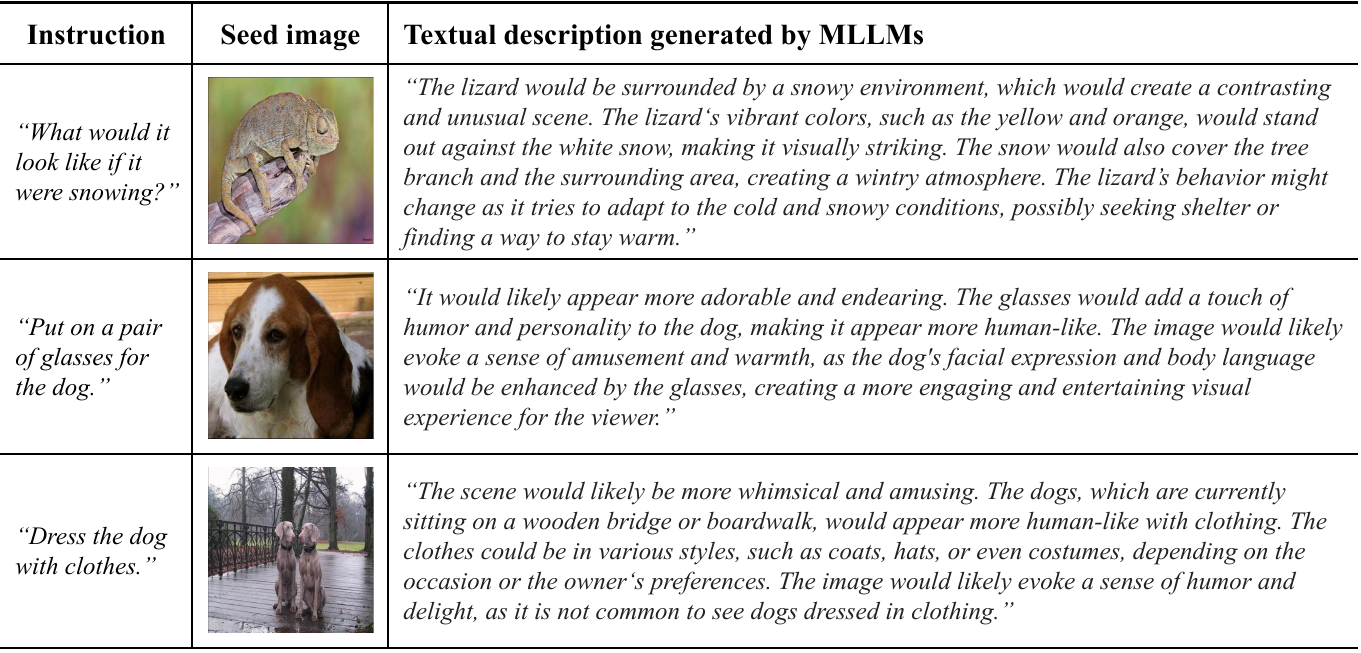}
  \caption{Textual descriptions generated in the bidirectional \& post-hoc
  pipeline of MLLM-based mutations.}
  \label{fig:new-text}
\end{figure*}

\F~\ref{fig:new-text} shows examples of textual descriptions in the
bidirectional pipeline. The text can precisely describe the mutation expected
by the instruction. For instance, in the first case shown in
\F~\ref{fig:new-text}, the description \textit{``The lizard would be surrounded
by a snowy environment...''} precisely captures the semantic transformation
specified by the mutation instruction \textit{``What would it look like if it
were snowing''}, facilitating accurate execution of the intended mutation. In
addition, the text includes many detailed and informative descriptions, e.g.,
\textit{``The snow would also cover the tree branch and the surrounding
area...''} in the first case, which gives more information of the mutated image
to ease the mutant generation.

Overall, the bidirectional pipeline, to some extent, achieves automated prompt
engineering by enriching the mutation instruction with the seed image's details.
The unidirectional pipeline, on the other hand, fine-tuning the Diffusion model
to strengthen its capability in achieving image mutations.
In our evaluations, we incorporate three MLLMs, namely, \vlm, \llava, and \gpt,
into the bidirectional pipeline, resulting in three MLLM-based mutation methods.
We use the unidirectional pipeline with its tailored MLLM and denote this
mutation method as \insp.
It's worth noting that \vlm, \llava, \gpt, and \insp\ (before fine-tuning)
use the same Stable Diffusion model~\cite{rombach2022high} as the
image generation module, ensuring a fair comparison for the capabilities of
different MLLMs.

\subsection{Datasets and Mutations}
\label{subsec:dataset}

\T~\ref{tab:dataset} lists the evaluated datasets. They are large-scale and
representative to cover different application scenarios. ImageNet consists of
1,000 classes of real-life images, and is widely used for general image
classification tasks. Dog-Breed dataset contains dog images with 120 different
breeds and is a subset of ImageNet. It is constructed to improve VDL model's
accuracy in domain-specific applications. FFHQ includes human face photos; it is
widely used for face recognition and annotation tasks. CityScape is composed of
urban street scene images, which are popular in autonomous driving, object
detection, and instance segmentation, etc.

\begin{table*}[!htbp]
  \centering
  \caption{Evaluated datasets.}
  \label{tab:dataset}
  \resizebox{0.8\linewidth}{!}{
    \begin{tabular}{l|c|c|c}
      %@{\hspace{2pt}}c@{\hspace{2pt}}|
      %@{\hspace{2pt}}c@{\hspace{2pt}}|
      %@{\hspace{2pt}}c@{\hspace{2pt}}|
      %@{\hspace{2pt}}c@{\hspace{2pt}}
      \hline
      \textbf{Dataset}                                       & \textbf{\#Classes}   & \textbf{Description}                       & \textbf{VDL Tasks}                              \\
      \hline
      ImageNet~\cite{deng2009imagenet}                       & 1,000                & Real-life images                   & Image Classification                            \\
      \hline
      Dog-Breed~\cite{dogbreed}                              & 120                  & Dog images                  & Image Classification                            \\
      \hline
      FFHQ~\cite{karras2019style}           & N/A & Face photos         & Face recognition, annotation, etc. \\
      \hline
      \multirow{2}{*}{CityScape~\cite{cordts2016cityscapes}} & \multirow{2}{*}{N/A} & \multirow{2}{*}{Urban street scene images} & \multirow{2}{*}{\shortstack{Autonomous driving, object detection, \\ instance segmentation, etc.}} \\
                                                             &                      &                                            &                                                 \\
      \hline
      
    \end{tabular}
  }
\end{table*}

\begin{table*}[!htbp]
  \centering
  \caption{Mutations and their text instructions in MLLMs.}
  \label{tab:mutation}
  \resizebox{0.8\linewidth}{!}{
    \begin{tabular}{l|c|c}
      %@{\hspace{1pt}}l@{\hspace{1pt}}|
      %@{\hspace{2pt}}c@{\hspace{2pt}}|
      %@{\hspace{1pt}}c@{\hspace{1pt}}}
      % \begin{tabular}{c|l|c|c|c|Sc}
      \hline
      \textbf{Mutation}        & \textbf{Type}                         & \textbf{Instruction}                                           \\
      \hline
      \ding{172}~\pixel\       & Pixel                                 & ``\textit{Increase the contrast of the image.}''               \\
      \hline
      \ding{173}~\rotate\      & Geo.                           & ``\textit{Rotate the image 180 degrees.}''                     \\
      \hline
      \ding{174}~\style\       & \multirow{3}{*}{Style}                & ``\textit{Change the style of this image to Van Gogh style.}'' \\
      \ding{175}~\weather\     &                                       & ``\textit{What would it look like if it were snowing?}''(for general images)       \\
      \ding{176}~\driving\     &                                       & ``\textit{What would it look like if it were snowing?}''(for driving scene photos)       \\
      \hline
      \ding{177}~\face\        & Percep.                            & ``\textit{Make the eyes of the person close.}''                \\
      \hline
      \ding{178}~\background\  & \multirow{4}{*}{\shortstack{MLLM-\\enabled}} & ``\textit{Change the background to a library.}''   \\
      \ding{179}~\clothes\     &                                       & ``\textit{Dress the dog with clothes.}''                       \\
      \ding{180}~\legs\        &                                       & ``\textit{Change the dog's legs to be bionic.}''               \\
      \ding{181}~\glasses\     &                                       & ``\textit{Put on a pair of glasses for the dog.}''             \\
      \hline
    \end{tabular}
  }
  \begin{tablenotes}
    \item \weather: weather transfer mutations applied on general images.
    \driving: weather transfer mutations applied on driving scene photos.
  \end{tablenotes}
  % \vspace{-5pt}
\end{table*}

\parh{Traditional.}~We consider five representative mutations for the five
categories of traditional mutations introduced in \S~\ref{sec:mutation}.
Additionally, since the weather transfer mutation is affected by the
to-be-mutated image, we evaluate it on both general images (\weather\ in
\T~\ref{tab:mutation}) and driving scene photos (\driving\ in
\T~\ref{tab:mutation}) from ImageNet and CityScape, respectively. The
perceptual-level mutation is achieved using GANs~\cite{zhu2021low}; it is
dataset-specific and only applies to aligned face photos.

\parh{MLLM-Enabled.}~For the semantic-replacement mutations uniquely enabled by
MLLMs, we consider four different variants operating in different granularities
and manifesting varying characteristics. \background\ only edits background
semantics, while the remaining three focus on finer-grained semantics in the
main subject. \glasses\ is more ``silent'': it replaces the original eyes with
glasses + new eyes, and the new eyes should be mostly similar to the original
ones, with only slight updates on the lighting condition for coherence.
\clothes\ and \legs, in contrast, are more active by \textit{completely}
replacing target semantics with new ones, and their target semantics are of
interest to different VDL models when recognizing the dog.
Since these four mutations require specifying the mutated object in the images,
to evaluate them uniformly on diverse images, we use the Dog-Breed dataset. That
said, extending them to other objects/images should be straightforward: users
only need to replace the word ``\texttt{dog}'' with the target one.

\parh{Prompt.}~To implement a mutation (either traditional or MLLM-enabled)
using MLLMs, a text instruction (i.e., prompt) is required. We list the text
instruction of each mutation in \T~\ref{tab:mutation}. Recall as mentioned in
\S~\ref{subsec:llm-pipeline}, we consider two optimized MLLM pipelines that are
specifically designed for achieving image mutations, so that a mutation's prompt
can be a single-sentence instruction, enabling an out-of-the-box usage. Due to
such a simple form, we find that the MLLM is mostly robust to changes over those
instructions, and we observe that, the instruction's wording, grammatical
structure, etc., does \textit{not} notably impact the achieved mutation.
We thus use the same instructions for both MLLM pipelines in our evaluations;
this setup also allows us to evaluate the specialties of different pipelines
and motivates future pipeline design.

\begin{table*}[!htbp]
  % \vspace{-5pt}
  \centering
  \caption{Results of human evaluations. We mark results within $[0, 1)$, $(2, 4]$, and $(4, 5]$
    in \colorbox{red!25}{red}, \colorbox{blue!25}{blue}, and \colorbox{green!25}{green}, respectively.}
  \label{tab:human-eval}
  \resizebox{0.95\linewidth}{!}{
    %  \begin{threeparttable}
    \begin{tabular}{c|c|c|c|c|c|c|c|c|c|c|c}
      \hline
      \textbf{Metric}               &             & \pixel\                  & \rotate\                 & \style\                  & \weather\                & \driving\                & \face\                   & \background\            & \clothes\                & \legs\                   & \glasses\                \\
      \hline
      \hline
      \multirow{5}{*}{Alignment}    & \llava\     & \colorbox{red!25}{0.8}   & \colorbox{red!25}{0.0}   & \colorbox{blue!25}{3.2}  & \colorbox{blue!25}{2.3}  & \colorbox{green!25}{4.3} & 1.3                      & 1.3                     & 1.9                      & 1.8                      & \colorbox{blue!25}{3.8}  \\
                                    & \vlm\       & \colorbox{red!25}{0.6}   & \colorbox{red!25}{0.1}   & \colorbox{blue!25}{3.1}  & 2.0                      & \colorbox{blue!25}{3.3}  & \colorbox{red!25}{0.7}   & 1.1                     & 1.4                      & 1.3                      & \colorbox{blue!25}{2.4}  \\
                                    & \gpt\       & \colorbox{red!25}{0.7}   & \colorbox{red!25}{0.1}   & \colorbox{blue!25}{3.1}  & \colorbox{blue!25}{2.1}  & \colorbox{blue!25}{3.8}  & 1.8                      & 1.3                     & \colorbox{blue!25}{2.3}  & \colorbox{blue!25}{2.1}  & \colorbox{blue!25}{3.2}  \\
                                    & \insp\      & 1.2                      & \colorbox{red!25}{0.0}   & \colorbox{blue!25}{2.5}  & \colorbox{red!25}{0.6}   & \colorbox{red!25}{0.1}   & \colorbox{red!25}{0.1}   & \colorbox{blue!25}{2.6} & \colorbox{red!25}{0.7}   & 1.1                      & \colorbox{green!25}{4.2} \\ \cline{2-12}
                                    & Traditional & \colorbox{green!25}{4.9} & \colorbox{green!25}{4.9} & 1.7                      & \colorbox{red!25}{0.1}   & \colorbox{green!25}{4.2} & \colorbox{green!25}{4.1} & N/A                     & N/A                      & N/A                      & N/A                      \\
      \hline
      \hline
      \multirow{5}{*}{Faithfulness} & \llava\     & \colorbox{blue!25}{3.0}  & \colorbox{blue!25}{3.6}  & \colorbox{blue!25}{3.3}  & \colorbox{blue!25}{2.5}  & 2.0                      & \colorbox{blue!25}{3.3}  & \colorbox{blue!25}{2.9} & \colorbox{blue!25}{3.0}  & \colorbox{blue!25}{3.0}  & \colorbox{blue!25}{3.5}  \\
                                    & \vlm\       & \colorbox{blue!25}{2.7}  & \colorbox{blue!25}{3.5}  & \colorbox{blue!25}{3.0}  & \colorbox{blue!25}{2.3}  & 2.0                      & \colorbox{blue!25}{3.0}  & \colorbox{blue!25}{2.8} & \colorbox{blue!25}{2.5}  & \colorbox{blue!25}{3.1}  & \colorbox{blue!25}{3.3}  \\
                                    & \gpt\       & \colorbox{blue!25}{3.0}  & \colorbox{blue!25}{3.7}  & \colorbox{blue!25}{3.3}  & \colorbox{blue!25}{2.7}  & \colorbox{blue!25}{2.1}  & \colorbox{blue!25}{3.0}  & \colorbox{blue!25}{3.0} & \colorbox{blue!25}{2.8}  & \colorbox{blue!25}{3.1}  & \colorbox{blue!25}{3.5}  \\
                                    & \insp\      & \colorbox{green!25}{4.9} & 1.0                      & \colorbox{green!25}{4.2} & \colorbox{green!25}{4.6} & \colorbox{green!25}{5.0} & \colorbox{blue!25}{3.3}  & \colorbox{blue!25}{3.0} & \colorbox{green!25}{4.2} & \colorbox{green!25}{4.5} & \colorbox{green!25}{4.3} \\ \cline{2-12}
                                    & Traditional & \colorbox{green!25}{5.0} & \colorbox{green!25}{4.9} & \colorbox{green!25}{4.6} & \colorbox{green!25}{4.5} & \colorbox{red!25}{0.2}   & \colorbox{green!25}{4.4} & N/A                     & N/A                      & N/A                      & N/A                      \\
      \hline
      \hline
      \multirow{5}{*}{Validity}     & \llava\     & \colorbox{blue!25}{4.0}  & \colorbox{green!25}{4.5} & \colorbox{green!25}{4.1} & \colorbox{blue!25}{4.0}  & \colorbox{blue!25}{2.8}  & \colorbox{blue!25}{3.1}  & \colorbox{blue!25}{3.1} & \colorbox{blue!25}{3.3}  & \colorbox{blue!25}{3.6}  & \colorbox{green!25}{4.4} \\
                                    & \vlm\       & \colorbox{blue!25}{3.6}  & \colorbox{green!25}{4.3} & \colorbox{blue!25}{3.7}  & \colorbox{blue!25}{3.8}  & \colorbox{blue!25}{2.7}  & \colorbox{blue!25}{2.5}  & \colorbox{blue!25}{2.9} & \colorbox{blue!25}{3.2}  & \colorbox{blue!25}{3.6}  & \colorbox{blue!25}{4.0}  \\
                                    & \gpt\       & \colorbox{blue!25}{4.0}  & \colorbox{green!25}{4.2} & \colorbox{blue!25}{3.9}  & \colorbox{blue!25}{3.8}  & \colorbox{blue!25}{2.9}  & \colorbox{blue!25}{2.5}  & \colorbox{blue!25}{3.1} & \colorbox{blue!25}{3.3}  & \colorbox{blue!25}{3.8}  & \colorbox{green!25}{4.5} \\
                                    & \insp\      & \colorbox{green!25}{4.9} & \colorbox{red!25}{0.9}   & \colorbox{green!25}{4.1} & \colorbox{green!25}{4.8} & \colorbox{green!25}{4.9} & \colorbox{blue!25}{3.6}  & \colorbox{blue!25}{2.6} & \colorbox{green!25}{4.2} & \colorbox{green!25}{4.5} & \colorbox{green!25}{4.2} \\ \cline{2-12}
                                    & Traditional & \colorbox{green!25}{5.0} & \colorbox{green!25}{4.9} & \colorbox{green!25}{4.4} & \colorbox{green!25}{4.5} & \colorbox{blue!25}{2.5}  & \colorbox{green!25}{4.4} & N/A                     & N/A                      & N/A                      & N/A                      \\
      \hline
    \end{tabular}
  }
  % \begin{tablenotes}
  %   \footnotesize
  %   \item \colorbox{green!25}{Green}: Perfectly satisfy.
  %   \colorbox{blue!25}{Blue}: Satisfy.
  %   \colorbox{red!25}{Red}: Fail.
  % \end{tablenotes}
  % \vspace{-10pt}
\end{table*}

\section{Human Studies Setup}
\label{sec:human}

We form a group of 20 participants to evaluate the ten mutations in
\T~\ref{tab:mutation} according to the alignment, faithfulness, and validity
requirements (discussed in \S~\ref{subsec:aspects}).
Since evaluating if a mutation fulfills these requirements needs inspecting how
the changed semantics (both anticipated and unanticipated) may affect the
downstream VDL testing and analysis, we hire 20 Ph.D. students experienced in
software testing and VDL systems in our human study.
For each mutation in \T~\ref{tab:mutation}, we prepare 50 seed images to mutate.
That is, for the six traditional mutations, each of them has $50 \times 5 = 250$
mutated images, and for the four MLLM-enabled mutations, each of them has $50
  \times 4 = 200$ mutated images.

The common practice, as in most existing MLLM
works~\cite{zhang2024real,sheynin2023emu}, is letting humans rank different
mutated images. However, we observed that MLLMs can fail in implementing a
mutation or generating valid images, which cannot be reflected in the ranking
results. To address this, we design the questions as follows. Each question in
our human study is formed by a seed image and one of its mutated variants (using
different implementations), and a participant is asked to give a score (0 to 5)
for the \textit{alignment}, \textit{faithfulness}, and \textit{validity} of a
mutated image. A zero score indicates that the mutated image fails to satisfy
one requirement. For example, if the rotation mutation implemented using a MLLM
does not rotate an image, the alignment score should be zero. We use scores 1-5
to assess each requirement because one mutation has maximal five different
implementations (i.e., one traditional method + four MLLMs); the five different
choices should be distinguishable for different implementations.

We let each participant evaluate one of the ten mutations, as interchangeably
evaluating different mutations may confuse the participants, especially for
subtle mutations that require careful comparison. Thus, each participant answers
200 (for MLLM-enabled mutations) or 250 (for traditional mutations) questions.
To reduce personal preferences, each mutation is assigned to two different
participants. The orders of the questions are shuffled to avoid potential bias
due to question orders.

To rule out invalid evaluations (e.g., a participant misunderstands the
requirement), following previous work~\cite{yuan2022unveiling}, we
insert sanity-check questions into each evaluation without notifying the
participants. Specifically, we insert 10 questions where the ``mutated'' image
is identical to the seed image, and the participant is expected to assign 0, 5,
5 scores for the alignment, faithfulness, and validity requirements,
respectively. If a participant fails to pass 80\% of the sanity-check questions,
we will discard all his/her answered questions. It's worth noting that all our
participants pass the sanity check.

Before starting the human study, we prepare a 30-minute teaching for each
participant. The teaching primarily aims to introduce the three requirements
(i.e., alignment, faithfulness, and validity) with different successful and
failed examples. It also gives participants suggestions on giving scores. For
instance, zero indicates an unsatisfied requirement, and the scores should be
distinguishable. Our human study is conducted using the Amazon Mechanical Turk
platform~\cite{AMT}. We do not set a time limit for each question because we expect
participants to carefully assess the mutated images. The participant can pause
and resume the evaluation at any time to avoid fatigue. The average time to
finish all evaluation questions (i.e., 200 or 250) is around two hours
as reported by participants.

\section{Results and Findings.}
\label{sec:result}

This section presents the results of our human study and answers three RQs
mentioned in \S~\ref{sec:setup}.

\begin{figure*}[!htbp]
  \centering
  % \vspace{-5pt}
  \includegraphics[width=0.90\linewidth]{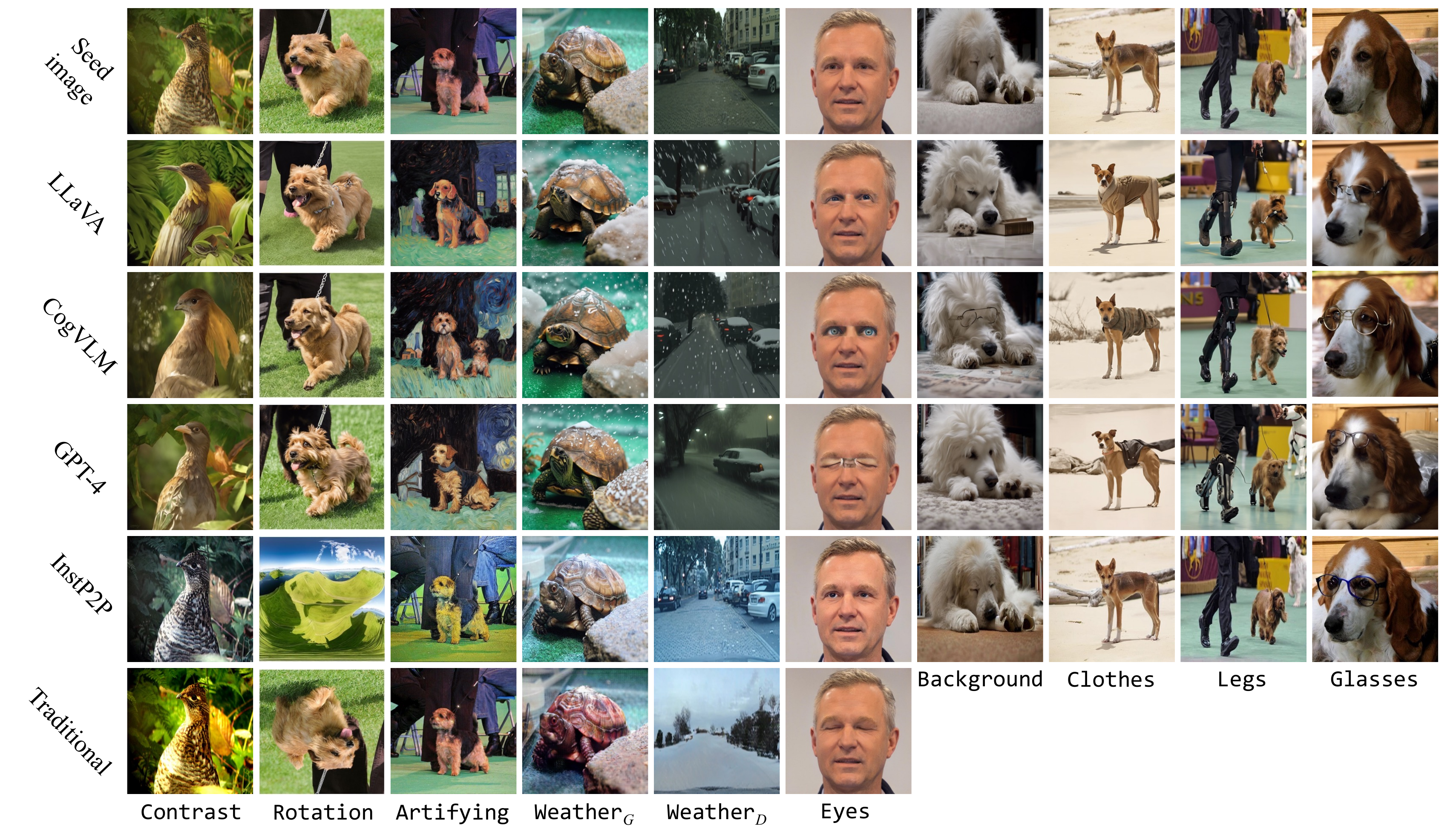}
  %\vspace{-5pt}
  \hspace{40pt}
  \caption{Mutation examples achieved via traditional method and different MLLMs.}
  %\vspace{-15pt}
  \label{fig:example}
\end{figure*}

\subsection{RQ1: Unifying and Generalizing Mutations}
\label{subsec:rq1}

\T~\ref{tab:human-eval} reports the average score over all 200 $\times$ 2 or 250
$\times$ 2 samples for each mutation. Since a zero score indicates that a
mutated image fails to satisfy one requirement, we deem scores within $[0, 1)$
as failing and mark them in red. We view scores larger than two as satisfying
the requirement. Specifically, scores within $(2, 4]$ are marked in blue,
whereas scores within $(4, 5]$ are marked in green to highlight the best method
for one mutation.

\F~\ref{fig:example} shows mutation examples of different methods. Below, we interpret our human
study results with these examples.

\parh{\pixel\ and \rotate.}~Our results show that MLLMs are unable to implement
pixel- and geometrical-level mutations, as supported by the red scores under the
alignment evaluation. Note that pixel- and geometrical-level mutations directly
operate on image pixels, whereas the remaining mutations focus on image features
(i.e., image representations obtained using neural networks). MLLMs also
primarily accept image features, leaving lower-level image semantics untouched.
This explains why MLLMs fail in implementing pixel- and geometrical-level
mutations, as images before and after these mutations often do not lead to
notable changes in image features. Recent MLLM papers have similar observations:
MLLMs perform worse in the spatial reasoning of image
contents~\cite{razeghi-etal-2022-impact,bang-etal-2023-multitask,cohn2023dialectical}.

\smallskip
Different from others, \insp\ fails in the validity evaluation of \rotate\ and
the faithfulness evaluation also has an average score of one. As shown in
\F~\ref{fig:example}, \insp\ generates a meaningless image when implementing
\rotate, and we find that similar failures occur in all evaluated samples.
Recall as noted in \S~\ref{subsec:llm-pipeline}, \insp\ differs from other MLLMs
by its unidirectional pipeline: to complement the missing details in the text
instruction, \insp\ fine-tunes the image generator (i.e., the Diffusion model) with
mutation samples. We therefore suspect that the fine-tuning harms the generative
capability of the Diffusion model. Hence, when implementing geometrical-level
mutations that require spatial reasoning, \insp\ fails to generate valid images.
In contrast, as other MLLMs do not modify the image generator, they mostly
``transmit'' the seed images if they cannot implement a mutation.

\parh{\style.}~MLLMs outperform traditional methods in transferring the artistic
style of images. Past methods often extract the artistic style from one or a few
target images, whose understanding of the target style may be limited. Moreover,
traditional style transfer may fail if contents in the seed image and the target
have distinct geometrical structures. These two limitations explain why the
alignment score of the traditional method is unsatisfactory (i.e., only 1.7).
MLLMs encode general knowledge and can better comprehend the target style from
their immense amounts of training data.

However, MLLM's faithfulness results are relatively worse than those of
traditional methods. As shown in \F~\ref{fig:example}, besides transferring the
style, MLLMs also edit other irrelevant semantics in the mutated images. This
observation is consistent with our findings mentioned above: MLLMs often focus on
high-level features and neglect low-level details. As a result, such details are
lost and consequently ``inpainted'' with random ones when generating mutated
images.

\parh{\weather\ and \driving.}~\insp\ fails to transfer weather conditions for
general images and driving scenes. We attribute the reason to the lack of
fine-tuning data. To support transferring weather conditions, the
fine-tuning in \insp\ requires mutation variants of the seed image that contain
different weather conditions, and such data tuples are hard to prepare on
a large scale. Thus, the fine-tuned Diffusion model in \insp\ is unable to capture
the weather conditions, transmitting the input image to the output in most cases,
as indicated by the high faithfulness and validity score.
The traditional method leverages the cycle-consistency to achieve weather
transfer without requiring paired images as the guidance. This paradigm
successfully transfers the weather conditions for diving scenes, resulting in a
high alignment score. However, it fails to retain the faithfulness (i.e., a 0.2
score), such that mutated images do not preserve irrelevant semantics, as
presented in \F~\ref{fig:example}. In addition, the cycle-consistency relies on
the similar geometrical structure between the seed and target images; this
explains why traditional method is inapplicable to mutate weather conditions in
general images whose geometrical structures are largely distinct from driving
scenes.

The other three MLLMs, especially \gpt, have satisfactory alignment,
faithfulness, and validity when transferring weather conditions for both general
images and driving scenes. We attribute these to the general knowledge encoded in
MLLMs. Although fine-grained details are also modified in MLLM's generated
images (see \F~\ref{fig:example}), we envision MLLM's promising application in
testing VDL systems where weather conditions are crucial (e.g., auto-driving).

\parh{\face.}~Traditional methods outperform MLLMs when implementing
perceptual-level mutations. All MLLMs are unable to deliver satisfactory
alignment scores. As in \F~\ref{fig:example}, all MLLMs do not close the eyes
but replace them with new eyes in most cases. That is, these MLLMs understand
that the mutation modifying eyes but cannot precisely mutate eyes.
Moreover, recall that as mentioned in \S~\ref{sec:motivation}, the traditional
implementation of \face\ requires specifying the localization of eyes. To make
the comparison fair, we also provide the eye locations to MLLMs. However, as
shown in \F~\ref{fig:example}, MLLMs fail to generate smoothly mutated images and
the edited eyes look unnatural, leading to relatively lower validity scores.

\begin{tcolorbox}[size=small]
  \textbf{Answer to RQ1}: MLLMs cannot implement pixel- and geometrical-level
  mutations well, potentially due to the lost information of such low-level
  semantics when MLLMs extract image features. MLLMs are also unable to generalize
  perceptual-level mutations. However, MLLMs optimized with the bidirectional
  pipeline perform better in extending style-level mutations to more general
  scenarios. They also exhibit better resilience: since they do not modify the
  image generator, they will not generate invalid images in case they cannot
  achieve a mutation.

\end{tcolorbox}

\subsection{RQ2: Expanding Mutated Semantics}
\label{subsec:rq2}

\parh{MLLM-Enabled (\ding{178}-\ding{181}).}~As in \T~\ref{tab:human-eval}, we
find that only \insp\ can implement \background\ to a satisfactory level,
whereas the other three MLLMs tend not to edit the seed image in \background, as
indicated by the relatively high faithfulness and validity score (see examples
in \F~\ref{fig:example}). However, the trend is on the opposite when
implementing \clothes\ and \legs, where \insp\ has lower alignment scores but
much higher faithfulness and validity scores. For \glasses, all MLLMs have
satisfactory alignment, faithfulness, and validity scores, and \insp\
outperforms the other three MLLMs.

Overall, these results show that MLLMs work well in semantic-replacement
mutations, indicating MLLM's high potential in expanding prior mutations with a
new dimension. The varied results of different pipelines (see
\S~\ref{subsec:llm-pipeline}) reveal their specialties. Both optimized
pipelines support subtle mutations to the subjects. However, the unidirectional
pipeline is more suitable for background, while the bidirectional pipeline fits
better when largely mutating subjects in images.

\parh{Testing Effectiveness.}~To further study whether MLLM-enabled mutations
are useful in VDL testing, we compare their effectiveness in detecting VDL
failures with traditional mutations. \pixel\ (\ding{172}) and \rotate\
(\ding{173}) are implemented with their original methods. \style\ (\ding{174}) 
and \weather\ (\ding{175}) are implemented using MLLMs as MLLMs perform better
than their original implementations. Note that \weather\ denotes \driving's
application in general images; \driving\ is therefore omitted. We also omit
\face\ in this evaluation since \face\ (\ding{177}) is inapplicable to general
images and MLLM also cannot generalize it according to results in
\S~\ref{subsec:rq1}.

VDL-based classification is selected as one representative VDL application. We
choose three recent SoTA VDL-based classifiers, Vision Transformer
(ViT)~\cite{dosovitskiy2021an}, SwinTransformer (SwinT)~\cite{liu2021swin}, and
ConvNeXt~\cite{liu2022convnet}. These models are officially released by PyTorch
and are trained with ImageNet.

\begin{table*}[]
    \caption{Testing effectiveness of different mutations, i.e., averaged
    $(\#\text{VDL faults}) / (\#\text{test inputs})$.}
    \centering
    \label{tab:effectiveness}
    \scalebox{0.95}{
        \begin{tabular}{c|c|c|c|c|c|c|c|c|c}
            \hline
            \textbf{Model}                     &             & \pixel\                  & \rotate\                 & \style\                  & \weather\                                                & \background\            & \clothes\                & \legs\                   & \glasses\                \\ \hline
            \multirow{6}{*}{ConvNeXt}          & LLaVA       & \multirow{4}{*}{N/A$^{*}$}      & \multirow{4}{*}{N/A}        & 7.6\% & 4.8\% & 5.2\% & 7.7\% & 7.7\% & 2.8\% \\
                                               & CogVLM      &       &         & 9.6\% & 4.8\% & 7.6\% & 8.6\% & 10.1\% & 4.5\% \\
                                               & GPT4V       &       &         & 9.2\% & 3.6\% & 6.8\% & 5.7\% & 9.7\% & 1.6\% \\
                                               & InstructP2P &       &         & 6.4\% & 4.4\% & 11.2\% & 0.8\% & 1.2\% & 1.2\% \\ \cline{2-10}
                                               & Traditional & 6.0\% & 0.4\% & 6.8\% & 14.4\% & N/A & N/A & N/A & N/A \\ \hline
            \multirow{6}{*}{SwinTransformer}   & LLaVA       & \multirow{4}{*}{N/A}      & \multirow{4}{*}{N/A}        & 8.0\% & 4.8\% & 5.2\% & 7.7\% & 7.3\% & 2.8\% \\
                                               & CogVLM      &       &         & 9.6\% & 4.8\% & 7.2\% & 9.4\% & 10.9\% & 4.5\% \\
                                               & GPT4V       &       &         & 9.2\% & 4.4\% & 7.2\% & 6.1\% & 9.3\% & 1.6\% \\
                                               & InstructP2P &       &         & 5.6\% & 4.0\% & 13.2\% & 2.0\% & 2.8\% & 1.6\% \\ \cline{2-10}
                                               & Traditional & 6.8\% & 0.4\% & 8.4\% & 16.0\% & N/A & N/A & N/A & N/A \\ \hline
            \multirow{6}{*}{VisionTransformer} & LLaVA       & \multirow{4}{*}{N/A}      & \multirow{4}{*}{N/A}        & 8.0\% & 4.8\% & 4.4\% & 6.1\% & 7.7\% & 3.2\% \\
                                               & CogVLM      &       &         & 9.6\% & 5.2\% & 7.6\% & 9.8\% & 10.5\% & 4.9\% \\
                                               & GPT4V       &       &         & 7.6\% & 4.4\% & 7.6\% & 5.3\% & 10.1\% & 0.8\% \\
                                               & InstructP2P &       &         & 7.2\% & 3.6\% & 12.0\% & 1.6\% & 2.0\% & 1.6\% \\ \cline{2-10}
                                               & Traditional & 8.4\% & 0.4\% & 7.2\% & 16.0\% & N/A & N/A & N/A & N/A \\ \hline
        \end{tabular}
      }
    \begin{tablenotes}
    \footnotesize
    \item* N/A indicates that the mutation cannot be implemented using MLLMs.
  \end{tablenotes}
\end{table*}

Mutated images are fed into a popular tool DeepHunter~\cite{xie2019deephunter}
to detect VDL faults. Since different mutations may change the seed image to
different extents, to avoid this, DeepHunter implements a filter to ensure that
the number of changed pixels and the modifications of pixel values in all test
inputs are at the same level. Results of the averaged $(\#\text{VDL faults}) /
  (\#\text{test inputs})$ are in \T~\ref{tab:effectiveness}. We see that the four
semantic-replacement mutations (\ding{178}-\ding{181}) are effective and
comparable to traditional mutations, indicating MLLM-enabled mutations are
valuable complements to existing mutations in VDL testing.

\begin{tcolorbox}[size=small]
  \textbf{Answer to RQ2}: MLLMs are complementary to traditional mutations and
  expand their mutated semantics by uniquely enabling new semantic-replacement
  mutations. These mutations are also effective in detecting VDL faults. The
  unidirectional mutation pipeline of MLLMs performs better for background
  semantics, whereas the bidirectional pipeline is suitable for noticeable
  mutations on subject semantics.
\end{tcolorbox}

\subsection{RQ3: Validation Metrics}
\label{subsec:rq3}

Aligned to the three aspects considered in our human study, we evaluate whether
existing validation metrics can reflect the validity of test inputs generated by
MLLM-enabled mutations and capture the varied extents of MLLM-enabled mutations.

\subsubsection{Validity Metrics}

Following existing works~\cite{harel2020neuron,yuan2023revisiting}, we take two popular metrics, Inception Score
(IS)~\cite{salimans2016improved} and Frechet Inception Distance
(FID)~\cite{heusel2017gans}, to measure the validity of the mutated images. Both
IS and FID use a set of real images as the reference, and compare the mutated
images with these reference images to estimate how likely the mutated images are
real. Lower FID values indicates better realism, whereas higher IS values
indicates better realism.

\begin{table}[!htbp]
  % \vspace{-5pt}
  \caption{Validity results of MLLM-enabled mutations.}
  \label{tab:val-llm}
  %\vspace{-5pt}
  \centering
  \resizebox{0.75\linewidth}{!}{
    \begin{tabular}{l|c|c|c|c}
      %@{\hspace{2pt}}c@{\hspace{2pt}}|
      %@{\hspace{2pt}}c@{\hspace{2pt}}|
      %@{\hspace{2pt}}c@{\hspace{2pt}}|
      %@{\hspace{2pt}}c@{\hspace{2pt}}|
      %@{\hspace{2pt}}c@{\hspace{2pt}}}
      \hline
          & \llava\  & \vlm\  & \gpt\  & \insp\ \\ \hline
      IS  & 14.51    & 13.71  & 14.82  & 23.93  \\ \hline
      FID & 88.09    & 89.65  & 88.67  & 93.95  \\ \hline
    \end{tabular}}
  %\vspace{-5pt}
\end{table}

As shown in \T~\ref{tab:val-llm}, IS values can roughly distinguish \insp\
from other MLLMs. While this is aligned with our human evaluation results, we expect
finer-grained differences among different MLLMs can be accurately reflected.
FID values, however, show contradictory results with IS values. Overall,
observations in \T~\ref{tab:val-llm} indicate the inaccuracy of existing
validity metrics when facing MLLM-enabled mutations.
As pointed out by previous
works~\cite{sajjadi2018assessing,kynkaanniemi2019improved}, FID and IS jointly
consider the validity and diversity. This was not an issue previously because
seed and mutated images are highly similar when applying most traditional
mutations (e.g., \rotate). However, semantic-replacement mutations enabled by
MLLMs bring non-determinism: multiple possible mutants can be generated when
applying one mutation to a seed image. As a result, the validity metric's
results can be falsely improved by diversified mutants in MLLM-enabled
mutations.

\subsubsection{Embedding-Based Metrics}

\begin{table}[!htbp]
  \centering
  %\vspace{-5pt}
  \caption{Anticipated results of MLLM-enabled mutations.}
  \label{tab:caption}
  %\vspace{-5pt}
  \resizebox{1.0\linewidth}{!}{
    \begin{tabular}{c|c}
      \hline
      \textbf{Mutation} & \textbf{Textual description of mutation results}        \\ \hline
      \background\      & ``\textit{The background of this image is a library.}'' \\ \hline
      \clothes\         & ``\textit{A dog wearing clothes.}''                     \\ \hline
      \legs\            & ``\textit{A dog with bionic legs.}''                    \\ \hline
      \glasses\         & ``\textit{A dog with glasses.}''                        \\ \hline
    \end{tabular}
  }
  % \vspace{-5pt}
\end{table}

As mentioned in \S~\ref{subsec:implicit}, for mutations lacking explicit formulas
(i.e., style-level, perceptual-level, and MLLM-enabled mutations), existing
works propose to measure them via image embeddings.
Following recent MLLM
papers~\cite{sheynin2023emu,zhang2024real,hertz2022prompt,missaoui2023semantic},
we obtain image embeddings via the SoTA embedding model
CLIP~\cite{radford2021learning}. The CLIP can obtain unified embeddings for
image semantics and text descriptions, enabling their similarity measurement. We
first prepare the textual description of the anticipated mutation result for
each mutation, as shown in \T~\ref{tab:caption}. For example, when mutating a
dog image with \glasses, the mutated image should have ``a dog with glasses''.
We then use the CLIP to extract the embedding vectors for the mutated image and
its corresponding textual description in \T~\ref{tab:caption}; their cosine
similarity is adopted to quantify the alignment of the mutation.

The faithfulness evaluation requires separating the local and fine-grained
semantics such as eyes and legs from the image, which is challenging especially
when the locations and postures of the dog are diverse. Therefore, we design a
coarse-grained metric that only considers an image's background and subject.
Specifically, following~\cite{hua2023dreamtuner}, we employ the SoTA subject
extractor InSPyReNet~\cite{kim2022revisiting} to separate the subjects and
background for the seed and mutated images. For \background, we use the
similarity of the two subjects in the seed and mutated image as the faithfulness
result, whereas for \ding{179}-\ding{181}, we use the similarity of two
backgrounds for faithfulness. Since semantic-replacement mutations may alter the
subject's outline (e.g., adding clothes may change the dog's shape), we compute
the similarity as the cosine similarity between CLIP image embeddings of the
extracted subjects/backgrounds.

\begin{table}[!htbp]
  %\vspace{-2pt}
  \caption{Alignment (left) and faithfulness (right) results of MLLM-enabled mutations.}
  \label{tab:eval-llm}
  %\vspace{-5pt}
  \centering
  \resizebox{1.0\linewidth}{!}{
    \begin{tabular}{c|cc|cc|cc|cc}
      \hline
               & \multicolumn{2}{c|}{\background} & \multicolumn{2}{c|}{\clothes} & \multicolumn{2}{c|}{\legs} & \multicolumn{2}{c}{\glasses}                             \\ \hline
      \llava\  & 0.67                             & 0.96                          & 0.49                       & 0.97                         & 0.43 & 0.97 & 0.50 & 0.96 \\ \hline
      \vlm\    & 0.67                             & 0.96                          & 0.49                       & 0.97                         & 0.43 & 0.96 & 0.50 & 0.96 \\ \hline
      \gpt\    & 0.67                             & 0.96                          & 0.49                       & 0.97                         & 0.43 & 0.97 & 0.50 & 0.97 \\ \hline
      \insp\   & 0.69                             & 0.95                          & 0.50                       & 0.97                         & 0.44 & 0.97 & 0.52 & 0.97 \\ \hline
    \end{tabular}
  }
\end{table}

\T~\ref{tab:eval-llm} presents the results of alignment and faithfulness
evaluations for four MLLM-enabled mutations using popular embedding-based
metrics. The alignment results of \background\ are relatively better than that of
\clothes, \legs, and \glasses, indicating that the metric can detect the
mutated background, but is unable to detect the mutated finer-grained semantics
on subjects. The high faithfulness results show that the embedding-based
metric can reflect the (ought-to-be) unchanged semantics.
Nevertheless, unlike our human studies that show notable differences in
alignment and faithfulness results of different MLLMs, the two numerical metrics
cannot distinguish their varied results, despite that similar methods show
promising results for traditional mutations~\cite{missaoui2023semantic}.

Note that traditional mutations like style- and perceptual-level mutations only
apply to specific domains, and accordingly, their corresponding image embeddings
can be limited to the target domain to improve precision. E.g., when measuring
mutations applied to face photos, existing works leverage face generators to
obtain face embeddings and do not consider non-face images. Similarly,
when measuring the achieved style-level mutations, prior works leverage style
extractor to specifically compute the style distance. However, MLLM-enabled
mutations are applicable to more general images. Although CLIP can obtain
embeddings for diverse images from different domains, these embeddings
consequently sacrifice their precision; similar observations are also noted in
recent research~\cite{yao2021filip,zhong2022regionclip}.

\begin{tcolorbox}[size=small]
  \textbf{Answer to RQ3}: Prior validation metrics are less applicable to
  MLLM-enabled mutations. Validity metrics like IS and FID jointly consider
  validity and diversity, whose numerical results can be misled by more
  diversified test inputs generated via MLLMs. Adapting metrics that separate
  validity and diversity (e.g., the Precision \& Recall
  metric~\cite{sajjadi2018assessing,kynkaanniemi2019improved}) for MLLMs may be
  a good starting point.
  Embedding-based metrics also become inaccurate due to the broader application
  scopes of MLLM-enabled mutations. Considering the high challenge of designing
  unified and accurate embedding models for diverse images, leveraging MLLMs to
  generate textual descriptions for mutants (following the bidirectional
  pipeline in \S~\ref{subsec:llm-pipeline}) may be promising to measure the
  achieved extents of MLLM-enabled mutations.

\end{tcolorbox}
\section{Discussion and Threat To Validity}
\label{sec:discussion}

This paper aims to understand the status quo of MLLM-based VDL testing and
provide insights. We also explicate the further efforts required to link MLLMs with
VDL testing. Our study enhances the confidence in using MLLMs (and in general,
LLMs) within the testing community. While to date, one can expect an
``out-of-the-box'' usage of MLLMs for testing VDL systems under some real-world
scenarios, we present the following lessons and discussions.

\parh{New Mutations.}~Our evaluation shows that MLLMs can generalize mutations
\textit{transferring semantics} between images, e.g., style-level mutations that
transfer styles, and enable semantic-replacement mutations that add new
semantics (from other images) to the seed. These are highly valuable in VDL
testing, as they can generate diverse and valid test inputs, and importantly,
these mutations are largely limited or hardly feasible with traditional methods.
We thus believe it is high time to incorporate MLLMs into VDL testing tasks,
which shall enable comprehensive and finer-grained testing of VDL systems
where subtle semantic changes are critical and can lead to
catastrophic failures, such as in autonomous driving and drone navigation.

Nevertheless, contemporary MLLMs are less capable of \textit{editing existing
semantics}. That said, despite the optimism, MLLMs are not a panacea for all VDL
testing tasks. We thus advocate to integrate and carefully choose
MLLMs and traditional methods in VDL testing.

\parh{Threat to Validity.}~The human study may be biased, and to mitigate this threat, we have involved 20
participants with relevant expertise, and each question is evaluated by
different participants to reduce personal bias. We also conduct a sanity check
to rule out answers from participants who likely misunderstand the questions. We
presume that it is sufficient to guarantee the credibility of our study.

Second, we also study the specialties of different pipelines of employing MLLMs
for mutations. All two available pipelines are considered in our evaluations,
and we incorporate four SoTA MLLMs into them to implement 10 representative
mutations. We believe the evaluated cases are comprehensive to alleviate potential
bias.

Third, the suitability of existing validation metrics for MLLM-based mutations
are also evaluated. To make the selected metrics representative, we consider
metrics that are widely adopted and have been proven effective in existing testing
works. We believe this setup is sufficient to reflect the common issues in validation
metrics tailored for traditional mutations.

\section{Conclusion}
\label{sec:conclusion}

This paper presents an in-depth study on MLLM-based mutations in VDL testing
and reveals the complementary positioning of MLLMs and traditional mutations.
We provide insights into the (in-)adequacy of MLLMs for VDL
testing, the specialties of different mutation pipelines of MLLMs, and the limitations
of existing validation metrics when facing MLLM-based mutations. 
Our findings call for careful integration of MLLMs and traditional image
mutations for better testing effectiveness and application scenarios.

\bibliographystyle{IEEEtran}
\bibliography{bib/main}

\end{document}